\documentclass[pra,aps,twocolumn,floatfix,showpacs]{revtex4}
\usepackage{graphicx,amsmath,amssymb,times}

\begin{document}
\title{Strong-coupling perturbation theory for the extended Bose-Hubbard model}
\author{M. Iskin$^1$ and J. K. Freericks$^2$}
\affiliation{$^1$Joint Quantum Institute, National Institute of Standards and Technology, and 
University of Maryland, Gaithersburg, MD 20899-8423, USA. \\
$^2$Department of Physics, Georgetown University, Washington, DC 20057, USA.}
\date{\today}

\begin{abstract}
We develop a strong-coupling perturbation theory for the extended Bose-Hubbard 
model with on-site and nearest-neighbor boson-boson repulsions on 
($d > 1$)-dimensional hypercubic lattices. Analytical expressions 
for the ground-state phase boundaries between the incompressible 
(Mott or charge-density-wave insulators) and the compressible 
(superfluid or supersolid) phases are derived up to third order in the 
hopping $t$. We also briefly discuss possible implications of our 
results in the context of ultracold dipolar Bose gases with dipole-dipole 
interactions loaded into optical lattices.

\end{abstract}

\pacs{03.75.Lm, 37.10.Jk, 67.85.-d}
\maketitle

\section{Introduction}
\label{sec:introduction}

Ultracold atomic physics in optical lattices has created a new experimental 
arena where many simple model Hamiltonians can be constructed and 
`simulated' experimentally~\cite{jaksch}. To date, the most successful 
efforts have been with bosonic atoms on optical lattices~\cite{greiner,spielman-1,spielman-2,bloch}. 
Here, when the single-particle bands of the optical lattice are well 
separated in energy, the boson-boson interaction is much smaller than 
that separation, and the particle filling is not too high, the system 
is described well by the single-band Bose-Hubbard (BH) model. This model 
is the bosonic generalization of the Hubbard model and was introduced originally 
to describe $^4$He in porous media or disordered granular superconductors~\cite{fisher}.  
The superfluid phase of bosonic systems is well described by weak-coupling 
theories, but the insulating phase, where there is a gap to 
particle excitations with a uniform (integer) filling of the bosons on 
each lattice site, is a strong-coupling phenomenon that only appears 
when the system is on a lattice. This Mott insulator phase is incompressible 
and hence occupies a finite area in the parameter space of the chemical 
potential and the hopping. It has a transition from the incompressible phase 
to a compressible superfluid as the hopping
or chemical potential are varied. The on-site BH model has been studied 
extensively, and the strong-coupling perturbation theory approach has been 
shown to be quite accurate in determining this phase diagram 
of the system~\cite{freericks-1,freericks-2}.

Recently, experimental progress has been made in constructing ultracold 
dipolar gases of molecules, namely K-Rb molecules, from a mixture of fermionic $^{40}$K and 
bosonic $^{87}$Rb atoms~\cite{ye-1,ye-2}. In this case, the molecules are fermionic, 
but similar principles will allow one to also create bosonic dipolar molecules. 
Future experiments are likely to load these bosonic molecules into optical lattices. 
These systems will have a long-range boson-boson interaction mediated by their dipole 
moment, which can be approximated, in some circumstances, by an on-site and 
a nearest-neighbor repulsion (generically, dipole-dipole interactions will be 
longer ranged than just nearest neighbors and also can have directionality 
due to the orientations of the dipoles). The case of an extended BH model, 
where the boson-boson interaction is longer ranged, has also been widely 
studied~\cite{bruder, parhat, otterlo, kuhner, kovrizhin}.
Inclusion of a nearest-neighbor repulsion can lead to the formation of 
a charge-density-wave (CDW) phase, where, at half-filling for example, 
one would have a checkerboard arrangement of the density 
in an ordered pattern. This phase is incompressible with a finite gap to excitations; 
it also breaks the original translational symmetry of the lattice, 
forming a new crystalline phase. The CDW phase has generated 
significant interest, because it often can become a supersolid
prior to becoming a superfluid as the interactions are reduced.  
A supersolid phase is a (compressible) superfluid that continues to have a 
density modulation (or CDW) present~\cite{leggett}; that is, the superfluid 
and crystalline orders co-exist. Interest in supersolid 
physics has increased dramatically since the recent observation of 
supersolid-like behavior in low-temperature He experiments~\cite{chan}.
There is some numerical and theoretical evidence that the supersolid phase 
exists only in dimensions higher than one~\cite{otterlo, kuhner}.

In this work, we examine the extended BH model with on-site and 
nearest-neighbor boson-boson interactions via a strong-coupling perturbation 
theory in the hopping, plus a scaling analysis, which allows us to accurately 
predict the critical point, and the shape of the insulating lobes in 
the plane of the chemical potential and the hopping.  We carry the analysis 
out to third-order in the hopping, and we perform the scaling theory using 
the known critical behavior at the tip of the insulating lobes 
[which corresponds to the ($d+1$)-dimensional $XY$ model, and is identical 
for the Mott and CDW phases].

The remainder of the manuscript is organized as follows. After introducing the
model Hamiltonian in Sec.~\ref{sec:ebh}, we develop the strong-coupling 
perturbation theory in the kinetic-energy term in Sec.~\ref{sec:sc},
where we derive analytical expressions for the phase boundaries between the 
incompressible (Mott or CDW insulators)
and compressible (superfluid or supersolid) phases. There we also propose a 
chemical-potential extrapolation technique based on scaling theory 
to extrapolate our third-order power series expansion into a functional 
form that is appropriate for the Mott or CDW lobes, and compare these results 
with the mean-field ones in Sec.~\ref{sec:mf}. 
A brief summary of our conclusions is presented in Sec.~\ref{sec:conclusions}.

\section{Extended Bose-Hubbard Model}
\label{sec:ebh}

We consider the following extended BH Hamiltonian with on-site and 
nearest-neighbor boson-boson repulsions
\begin{align}
\label{eqn:ebhh}
H = &- \sum_{i,j } t_{ij} b_i^\dagger b_j 
+ \frac{U}{2} \sum_i \widehat{n}_i (\widehat{n}_i-1) \nonumber \\
&+ \sum_{ i,j } V_{ij}\widehat{n}_i \widehat{n}_j -\mu \sum_i \widehat{n}_i,
\end{align}
where $t_{ij}$ is the tunneling (or hopping) matrix between sites $i$ and $j$, 
$b_i^\dagger$ ($b_i$) is the boson creation (annihilation) operator at site $i$,
$\widehat{n}_i = b_i^\dagger b_i$ is the boson number operator, 
$U>0$ is the strength of the on-site and 
$V_{ij}$ is the longer-ranged boson-boson repulsion between bosons at sites $i$ and $j$, 
and $\mu$ is the chemical potential.
In this manuscript, we assume $t_{ij}$ is a real symmetric matrix with elements
$t_{ij} = t$ for $i$ and $j$ nearest neighbors and $0$ otherwise and similarly 
for $V_{ij}$ (equal to $V>0$ for $i$ and $j$ nearest neighbors and zero otherwise), 
and consider a $(d > 1)$-dimensional hypercubic lattice with $M$ sites.   
Note that we work on a periodic lattice with no external trap potential.

We also assume $U > z V$ where $z=2d$ is the lattice coordination number 
(number of nearest neighbors). 
In this case, the boson occupancy of the nearest-neighbor sites in the 
CDW phase can only differ by one. For instance, the first CDW phase
is such that every other site is occupied by one boson and the remaining 
sites are left unoccupied. When $U <z V$, additional CDW phases can be 
present in the phase diagram. For instance, a CDW phase in which 
every other site is occupied by two bosons and the remaining sites 
are left unoccupied is energetically more favorable than a Mott phase 
in which every lattice site is occupied by one boson. 
Our results, with minor changes, can also be used to analyze 
these additional CDW phases if desired, but more work would be needed 
to examine other types of CDW order, like columnar (stripes) and so on, 
which can arise from longer-range interactions.

\subsection{The Atomic ($t = 0$) Limit}
\label{sec:al}

To understand the zero-temperature ($T = 0$) phase diagram of 
the extended BH model given in Eq.~(\ref{eqn:ebhh}), 
we start by analyzing the atomic ($t = 0$) limit. In this limit, since the 
kinetic energy vanishes, the boson number operator $\widehat{n}_i$ 
commutes with all of the remaining terms of the Hamiltonian.
Therefore, every lattice site is occupied by a fixed number 
$n_i$ of bosons and the system is insulating. 

When $V = 0$, the ground-state boson occupancy is the same for every 
lattice site such that $\langle \widehat{n}_i \rangle = n_0$ 
where $\langle ... \rangle$ is the thermal average, and the average boson occupancy
$n_0$ is chosen to minimize the ground-state energy for a given $\mu$ 
($n_0$ is an integer here and should not be confused with the condensate fraction of a superfluid). 
It turns out that the ground-state energy of the $n_0$ state is degenerate
with that of the $n_0+1$ state at $\mu = U n_0$. 
This means that the chemical potential width of all Mott lobes is $U$, and that 
the boson occupancy increases from $n_0$ to $n_0+1$ when $\mu = U n_0 + 0^+$.
For instance, the ground state is a vacuum with $n_0 = 0$ for $\mu \le 0$; 
it is a Mott insulator with $n_0 = 1$ for $0 \le \mu \le U$; it is a Mott insulator with $n_0 = 2$ 
for $U \le \mu \le 2U$, and so on.

When $V \ne 0$, the ground state has an additional CDW phase which
has crystalline order in the form of staggered boson densities, \textit{i.e.}
$\langle \widehat{n}_i \rangle = n_a$ and $\langle \widehat{n}_j \rangle = n_b$
for $i$ and $j$ nearest neighbors.  Therefore, to describe the CDW phases, 
it is convenient to split the entire lattice into two sublattices A and B such 
that the nearest-neighbor sites belong to a different sublattice 
(a lattice for which this can be done is called a bipartite lattice---
we assume the number of lattice sites in each sublattice is the same here). 
We assume that the boson occupancies of the sublattices A and B 
are $n_a$ and $n_b$, respectively, such that $n_a \ge n_b$. We remark that
the $n_a = n_b = n_0$ states correspond to the Mott phase.
It turns out that the ground-state energy of the $(n_a=n_0+1, n_b=n_0)$ state 
is degenerate with those of the $(n_a=n_0, n_b=n_0)$ and 
$(n_a=n_0+1, n_b=n_0+1)$ states at $\mu = U n_0 + z V n_0$ 
and $\mu = U n_0 + z V (n_0+1)$, respectively. This means that the chemical
potential width of all Mott and CDW lobes are $U$ and $z V$, respectively, 
and that the ground state alternates between the CDW and Mott phases as a 
function of increasing $\mu$. For instance, the ground state is a vacuum 
$(n_a=0, n_b=0)$ for $\mu \le 0$; it is a CDW with
$(n_a=1, n_b=0) $ for $0 \le \mu \le z V$; it is a Mott insulator with $(n_a=1, n_b=1)$ 
for $zV \le \mu \le U+zV$; it is a CDW with $(n_a=2, n_b=1)$ 
for $U+zV \le \mu \le U + 2zV$; it is a Mott insulator with $(n_a=2, n_b=2)$ for 
$U+2zV \le \mu \le 2U+2zV$, and so on.

Having discussed the $t = 0$ limit, now we are ready to analyze the 
competition between the kinetic and potential energy terms of 
the Hamiltonian when $t \ne 0$. 
As $t$ increases, one expects that the range of $\mu$ about which the 
ground state is insulating (incompressible) decreases, and that the Mott 
and CDW phases disappear at a critical value of $t$, beyond which 
the system becomes compressible.

\subsection{Transition from an Incompressible to a Compressible Phase}
\label{sec:transition}

To determine the phase boundary between the incompressible (Mott or CDW insulators) 
and the compressible (superfluid or supersolid) phases, we need the 
energies of the Mott and CDW phases and of their defect states 
as a function of $t$. The defect states are characterized by exactly one extra particle or 
hole which moves coherently throughout the lattice. At the point where the 
energy of the incompressible state becomes degenerate with its defect 
state, the system becomes compressible assuming that the compressibility
approaches zero continuously at the phase boundary. 
Therefore, the phase boundary between the Mott and superfluid phases 
is determined by
\begin{align}
\label{eqn:pb-mott}
E_{\rm Mott}^{\rm ins} (n_0) &= E_{\rm Mott}^{\rm par} (n_0), \\
E_{\rm Mott}^{\rm ins} (n_0) &= E_{\rm Mott}^{\rm hol} (n_0),
\end{align}
where $E_{\rm Mott}^{\rm ins} (n_0)$ is the energy of the Mott phase with $n_0$ 
bosons on every lattice site, and $E_{\rm Mott}^{\rm par} (n_0)$ and
$E_{\rm Mott}^{\rm hol} (n_0)$ are the energies of the Mott-defect phases
with exactly one extra particle or hole, respectively. These conditions determine
the phase boundaries of the particle and hole branches of the Mott insulating lobes, $\mu_{\rm Mott}^{\rm par}$ 
and $\mu_{\rm Mott}^{\rm hol}$, respectively, 
as a function of $t$, $U$, $V$ and $n_0$. Similarly the phase boundary
between the CDW and supersolid phases is determined by
\begin{align}
E_{\rm CDW}^{\rm ins} (n_a, n_b) &= E_{\rm CDW}^{\rm par} (n_a, n_b), \\
\label{eqn:pb-cdw}
E_{\rm CDW}^{\rm ins} (n_a, n_b) &= E_{\rm CDW}^{\rm hol} (n_a, n_b),
\end{align}
where $E_{\rm CDW}^{\rm ins} (n_a, n_b)$ is the energy of the CDW phase 
with $n_a$ and $n_b$ bosons on alternating lattice sites, and 
$E_{\rm CDW}^{\rm par} (n_a, n_b)$ and $E_{\rm CDW}^{\rm hol} (n_a, n_b)$ 
are the energies of the CDW-defect phases with exactly one extra particle or
hole, respectively. These conditions determine the phase boundaries of the 
particle and hole branches of the CDW insulating lobes, $\mu_{\rm CDW}^{\rm par}$ and $\mu_{\rm CDW}^{\rm hol}$, 
respectively, as a function of $t$, $U$, $V$, $n_a$ and $n_b$. 
Next, we calculate the energies of the Mott and CDW phases 
and of their defect states as a perturbative series in the hopping $t$.

\section{Strong-Coupling Perturbation Theory}
\label{sec:sc}

We use the many-body version of Rayleigh-Schr\"odinger 
perturbation theory in the kinetic energy term~\cite{landau} to perform
 the expansion (in powers of the hopping) for the different energies 
 needed to carry out our analysis. The perturbation theory is performed 
with respect to the ground state of the system when the kinetic-energy term 
is absent. This technique was previously used to discuss the phase diagram 
of the on-site BH model~\cite{freericks-1, freericks-2}, and its results showed an excellent 
agreement with the Quantum Monte Carlo simulations (including the most 
recent numerical work~\cite{prokofiev-1,prokofiev-2}). Here, we generalize 
this method to the extended BH model, hoping to develop an analytical approach 
which could also be as accurate as the numerical ones. However, we remark that our
strong-coupling perturbation theory cannot be used to calculate the phase 
boundary between two compressible phases, \textit{e.g.} the supersolid 
to superfluid transition.  In addition, we cannot even tell whether the 
compressible phase is a supersolid or a superfluid.

\subsection{Ground-State Wavefunctions at Zeroth Order in $t$}
\label{sec:wf}

For our purpose, we first need the ground-state wavefuntions of the 
Mott and CDW phases and of their particle and hole defects when $t = 0$.
To zeroth order in $t$, the Mott and CDW wavefunctions can be written as
\begin{eqnarray}
\label{eqn:wf-ins}
|\Psi_{\rm Mott}^{\rm ins (0)} \rangle &=& \prod_{k=1}^{M} \frac{(b_k^\dagger)^{n_0}}{\sqrt{n_0!}} | 0 \rangle, \\
|\Psi_{\rm CDW}^{\rm ins (0)} \rangle &=& \prod_{i \in A, j \in B}^{M/2} 
\frac{(b_i^\dagger)^{n_a}}{\sqrt{n_a!}} \frac{(b_j^\dagger)^{n_b}}{\sqrt{n_b!}} | 0 \rangle,
\end{eqnarray}
where $M$ is the number of lattice sites, and $| 0 \rangle$ is the vacuum state 
(here, we remind that the lattice is divided equally into A and B sublattices). 
We use, here and throughout, the index $k$ to refer to all lattice sites, 
while the indices $i$ and $j$ are limited to the A and B sublattices, respectively.

On the other hand, the wavefunctions of the defect states are determined 
by degenerate perturbation theory.  To zeroth order in $t$, the wavefunctions 
for the particle-defect states can be written as
\begin{eqnarray}
\label{eqn:wf-par}
|\Psi_{\rm Mott}^{\rm par (0)} \rangle &=& \frac{1}{\sqrt{n_0+1}}
\sum_{k=1}^{M} f_k^{\rm Mott} b_k^\dagger |\Psi_{\rm Mott}^{\rm ins (0)} \rangle, \\
|\Psi_{\rm CDW}^{\rm par (0)} \rangle &=& \frac{1}{\sqrt{n_b+1}}
\sum_{j \in B}^{M/2} f_j^{\rm CDWB} b_j^\dagger |\Psi_{\rm CDW}^{\rm ins (0)} \rangle,
\end{eqnarray}
where $f_k^{\rm Mott}$ is the eigenvector of the hopping matrix $t_{kk'}$ with the highest 
eigenvalue (which is $z t$) such that
$
\sum_{k'} t_{kk'} f_{k'}^{\rm Mott} = z t f_{k}^{\rm Mott},
$
and $f_j^{\rm CDWB}$ is the eigenvector of $\sum_{i} t_{ji} t_{ij'}$ 
(this matrix lives solely on the B sublattice) with the highest 
eigenvalue (which is $z^2 t^2)$ such that 
$
\sum_{ij'} t_{ji} t_{ij'} f_{j'}^{\rm CDWB} = z^2 t^2 f_{j}^{\rm CDWB}.
$
Notice that we choose the highest eigenvalue of $t_{ij}$ because the hopping matrix 
enters the Hamiltonian as $-t_{ij}$, and we ultimately want the lowest-energy states; 
similarly for the CDW phases, the coefficient of the $t^2$ matrix that enters the 
perturbation theory is negative, so we want the highest eigenvalue again.
The normalization condition requires that $\sum_{k=1}^{M} |f_k^{\rm Mott}|^2 =1$
and $\sum_{j \in B}^{M/2} |f_j^{\rm CDWB}|^2 =1$.
Similarly, to zeroth order in $t$, the wavefunctions for the hole-defect states can be written as
\begin{eqnarray}
\label{eqn:wf-hol}
|\Psi_{\rm Mott}^{\rm hol (0)} \rangle &=& \frac{1}{\sqrt{n_0}}
\sum_{k=1}^{M} f_k^{\rm Mott} b_k |\Psi_{\rm Mott}^{\rm ins (0)} \rangle, \\
|\Psi_{\rm CDW}^{\rm hol (0)} \rangle &=& \frac{1}{\sqrt{n_a}}
\sum_{i \in A}^{M/2} f_i^{\rm CDWA} b_i |\Psi_{\rm CDW}^{\rm ins (0)} \rangle,
\end{eqnarray}
where $f_i^{\rm CDWA}$ is the eigenvector of $\sum_{j} t_{ij} t_{ji'}$ 
(this matrix lives solely on the A sublattice) with the highest 
eigenvalue (which is $z^2 t^2)$ such that 
$
\sum_{j i'} t_{ij} t_{ji'} f_{i'}^{\rm CDWA} = z^2 t^2 f_{i}^{\rm CDWA}.
$
The normalization condition requires that $\sum_{i \in A}^{M/2} |f_i^{\rm CDWA}|^2 =1$.

\subsection{Ground-State Energies up to Third Order in $t$}
\label{sec:en}

Next, we employ the many-body version of Rayleigh-Schr\"odinger perturbation 
theory in $t$ with respect to the ground state of the system when $ t = 0$, and 
calculate the energies of the Mott and CDW phases and of their particle- and 
hole-defect states. To third order in $t$, the energy of the Mott state is obtained 
via nondegenerate perturbation theory and it is given by
\begin{align}
\label{eqn:en-mott}
\frac{E_{\rm Mott}^{\rm ins} (n_0)}{M} &= U \frac{n_0 (n_0-1)}{2} + z V \frac{n_0^2}{2} - \mu n_0  \nonumber \\ 
&-  n_0 (n_0+1) \frac{z t^2}{U-V} + O(t^4),
\end{align}
which is an extensive quantity, that is $E_{\rm Mott}^{\rm ins} (n_0)$ 
is proportional to the total number of lattice sites $M$. The odd-order terms in $t$ 
vanish for the $d$-dimensional hypercubic lattices considered in this manuscript. 
Notice that Eq.~(\ref{eqn:en-mott}) recovers the known result for the on-site BH model
when $V = 0$~\cite{freericks-1, freericks-2}. 
Similarly, to third order in $t$, the energy of the CDW state is also
obtained via nondegenerate perturbation theory and it can be written as
\begin{align}
\label{eqn:en-cdw}
&\frac{E_{\rm CDW}^{\rm ins} (n_a, n_b)}{M} = U \frac{n_a (n_a-1) + n_b(n_b-1)}{4} 
+ z V \frac{n_a n_b}{2} \nonumber \\
& - \mu \frac{n_a+n_b}{2}+ 
\left[ \frac{n_a (n_b+1)}{U(n_a-n_b-1)+V(z n_b-z n_a+1)} \right. \nonumber \\
&\left. + \frac{n_b (n_a+1)}{U(n_b-n_a-1)+V(z n_a-z n_b+1)} \right] \frac{z t^2}{2} + O(t^4),
\end{align}
which is also an extensive quantity, and the odd-order terms in $t$ also vanish. 
Notice that Eq.~(\ref{eqn:en-cdw}) reduces to Eq.~(\ref{eqn:en-mott}) 
when $n_a = n_b = n_0$ as expected.

The calculation of the defect state energies is more involved since it requires using
degenerate perturbation theory. This is because when exactly one extra particle or hole 
is added to the Mott phase, it could go to any of the $M$ lattice sites and all of 
those states share the same energy when $t=0$.
Therefore, for both Mott defect states with exactly one extra particle or hole, 
the initial degeneracy is of order $M$ and it is lifted at first order in $t$. 
A lengthy but straightforward calculation leads to the energy of the Mott 
particle-defect state up to third order in $t$ as
\begin{widetext}
\begin{align}
\label{eqn:en-mpar}
&E_{\rm Mott}^{\rm par} (n_0) = E_{\rm Mott}^{\rm ins} (n_0) + U n_0 + z V n_0 - \mu - (n_0+1) z t
+ n_0 \Big\lbrace 
(n_0+1) \left[ \frac{1-z}{U} + \frac{2(1-z)}{U-2V} + \frac{2z}{U-V} \right]
- \frac{n_0+2}{2(U-V)} \Big\rbrace z t^2 \nonumber \\
&- n_0 (n_0+1) \Big\lbrace 
n_0 \left[ \frac{z-2}{U^2} + \frac{z^2-3z+3}{(U-V)^2} \right] 
+ (n_0+1) \left[  \frac{z(1-z)}{U^2} - \frac{2z^2-6z+6}{(U-V)^2} + \frac{2z(1-z)}{(U-2V)^2} + \frac{2(z^2-3z+3)}{U(U-V)} \right. \nonumber \\
&\left. + \frac{4(z-2)}{U(U-2V)} + \frac{4(z^2-3z+3)}{(U-V)(U-2V)} \right] 
+ (n_0+2) \left[ \frac{z-1}{U(U-V)} - \frac{z}{4(U-V)^2} \right] 
\Big\rbrace z t^3 + O(t^4).
\end{align}
This expression is valid for all $d$-dimensional hypercubic lattices, and it
recovers the known result for the on-site BH model when $V = 0$~\cite{freericks-1, freericks-2}. 
To third order in $t$, we obtain a similar expression for the energy of the 
Mott hole-defect state given by
\begin{align}
\label{eqn:en-mhol}
&E_{\rm Mott}^{\rm hol} (n_0) = E_{\rm Mott}^{\rm ins} (n_0) - U (n_0-1) - z V n_0 + \mu - n_0 z t
+ (n_0+1) \Big\lbrace 
n_0 \left[ \frac{1-z}{U} + \frac{2(1-z)}{U-2V} + \frac{2z}{U-V} \right]
- \frac{n_0-1}{2(U-V)} \Big\rbrace z t^2 \nonumber \\
&- n_0 (n_0+1) \Big\lbrace 
(n_0+1) \left[ \frac{z-2}{U^2} + \frac{z^2-3z+3}{(U-V)^2} \right] 
+ n_0 \left[ \frac{z(1-z)}{U^2} - \frac{2z^2-6z+6}{(U-V)^2} + \frac{2z(1-z)}{(U-2V)^2} + \frac{2(z^2-3z+3)}{U(U-V)} \right. \nonumber \\
&\left. + \frac{4(z-2)}{U(U-2V)} + \frac{4(z^2-3z+3)}{(U-V)(U-2V)} \right]
+ (n_0-1) \left[ \frac{z-1}{U(U-V)} - \frac{z}{4(U-V)^2} \right] 
\Big\rbrace z t^3 + O(t^4),
\end{align}
which also is valid for all $d$-dimensional hypercubic lattices, and recovers 
the known result for the on-site BH model when $V = 0$~\cite{freericks-1, freericks-2}. 

On the other hand, for $d > 1$ dimensions, when an extra particle or hole is 
added to the CDW phase, it could go to any of the $M/2$ sites in the sublattice B or A, 
respectively (here, we remind that $n_a > n_b$ is assumed in this manuscript). 
Therefore, for both CDW defect states with an extra particle or hole in $d > 1$ 
dimensions, the degeneracy is of order $M/2$ and it is lifted at second order in $t$. 
This is because the states occupy one of the sublattices, and they cannot be connected 
by one hop, but rather require two hops to be connected.
Another lengthy but straightforward calculation leads to the energy of the CDW 
particle-defect state up to third order in $t$ as
\begin{align}
\label{eqn:en-cpar}
&E_{\rm CDW}^{\rm par} (n_a, n_b) = E_{\rm CDW}^{\rm ins} (n_a, n_b) + U n_b + z V n_a - \mu
+ \left[ 
\frac{(n_a+1)(n_b+1) z}{U(n_b-n_a)+V(z n_a-z n_b)} \right. \nonumber \\ 
&- \frac{n_a(n_b+1) z}{U(n_a-n_b-1)+V(z n_b-z n_a+1)} 
- \frac{n_b(n_a+1) z}{U(n_b-n_a-1)+V(z n_a-z n_b+1)} 
+ \frac{n_a(n_b+2)}{U(n_a-n_b-2)+V(z n_b-z n_a+2)} \nonumber \\
&\left.+ \frac{n_b(n_a+1) (z-1)}{U(n_b-n_a-1)+V(z n_a-z n_b)} + \frac{2n_a(n_b+1) (z-1)}{U(n_a-n_b-1)+V(z n_b-z n_a+2)}   
\right] z t^2 + O(t^4).
\end{align}
This expression is valid for $(d > 1)$-dimensional hypercubic lattices. 
Notice that the odd-order terms in $t$ vanish for these 
lattices. To third order in $t$, we obtain a similar
expression for the energy of the CDW hole-defect state given by
\begin{align}
\label{eqn:en-chole}
&E_{\rm CDW}^{\rm hol} (n_a, n_b) = E_{\rm CDW}^{\rm ins} (n_a, n_b) - U (n_a-1) - z V n_b + \mu
+ \left[ 
\frac{n_a n_b z}{U(n_b-n_a)+V(z n_a-z n_b)} \right. \nonumber \\
&- \frac{n_a(n_b+1) z}{U(n_a-n_b-1)+V(z n_b-z n_a+1)} 
- \frac{n_b(n_a+1) z}{U(n_b-n_a-1)+V(z n_a-z n_b+1)}
+ \frac{(n_a-1)(n_b+1)}{U(n_a-n_b-2)+V(z n_b-z n_a+2)} \nonumber \\
&\left. + \frac{n_b(n_a+1) (z-1)}{U(n_b-n_a-1)+V(z n_a-z n_b)} + \frac{2n_a(n_b+1) (z-1)}{U(n_a-n_b-1)+V(z n_b-z n_a+2)}   
\right] z t^2 + O(t^4).
\end{align}
\end{widetext}
This expression is also valid for $(d > 1)$-dimensional hypercubic lattices where the 
odd-order terms in $t$ vanish.

Notice that because the Mott defect states have corrections to first 
order in the hopping, while the CDW defects have corrections to second 
order in the hopping, the slopes of the Mott phase will be finite 
as $t\rightarrow 0$, but they will vanish for the CDW lobes.  
Hence, the shape of the different types of insulating lobes are always different.

In one dimension ($d = 1$), however, when exactly one extra particle or hole is 
added to the CDW phase, the degeneracy of both of the CDW defect states 
is of order $M$ and it is lifted at first order in $t$. This difference between 
$d > 1$ and $d = 1$ makes one dimension unique, and it is the reason that 
the supersolid phase exists in higher dimensions but not in one~\cite{otterlo, kuhner}. 
In other words, due to this large degeneracy, an extra particle or hole immediately 
delocalizes the bosons in $d = 1$, and the crystalline order disappears. 
Since $d=1$ requires special attention, it will be addressed elsewhere, 
and we restrict the analysis here to higher dimensions.

We would like to remark in passing that the energy difference between
the Mott and CDW phases with their defect states determine 
the phase boundary of the particle and hole branches. While all
$E_{\rm Mott}^{\rm ins} (n_0)$, $E_{\rm Mott}^{\rm par} (n_0)$ and 
$E_{\rm Mott}^{\rm hol} (n_0)$ depend on the lattice size $M$, their
difference does not. Therefore, the chemical potentials that determine 
the particle and hole branches, $\mu_{\rm Mott}^{\rm par}$ and 
$\mu_{\rm Mott}^{\rm hol}$, respectively, are independent of $M$ at the
phase boundaries. 
Similarly, while all $E_{\rm CDW}^{\rm ins} (n_a, n_b)$, 
$E_{\rm CDW}^{\rm par} (n_a, n_b)$ and $E_{\rm CDW}^{\rm hol} (n_a, n_b)$ 
depend also on the lattice size $M$, their difference does not. 
Therefore, the chemical potentials that determine the particle and hole branches, 
$\mu_{\rm CDW}^{\rm par}$ and $\mu_{\rm CDW}^{\rm hol}$, respectively,
are also independent of $M$ at the phase boundaries. 
These observations indicate that the numerical Quantum Monte Carlo 
simulations which are based on Eqs.~(\ref{eqn:pb-mott}) to (\ref{eqn:pb-cdw}) 
should not have too strong a dependence on $M$. It also shows that exact 
diagonalization on finite clusters of a sufficiently large size can also 
yield these expressions if properly analyzed to extract the coefficients 
of the power series.

\begin{figure} [htb]
\centerline{\scalebox{0.6}{\includegraphics{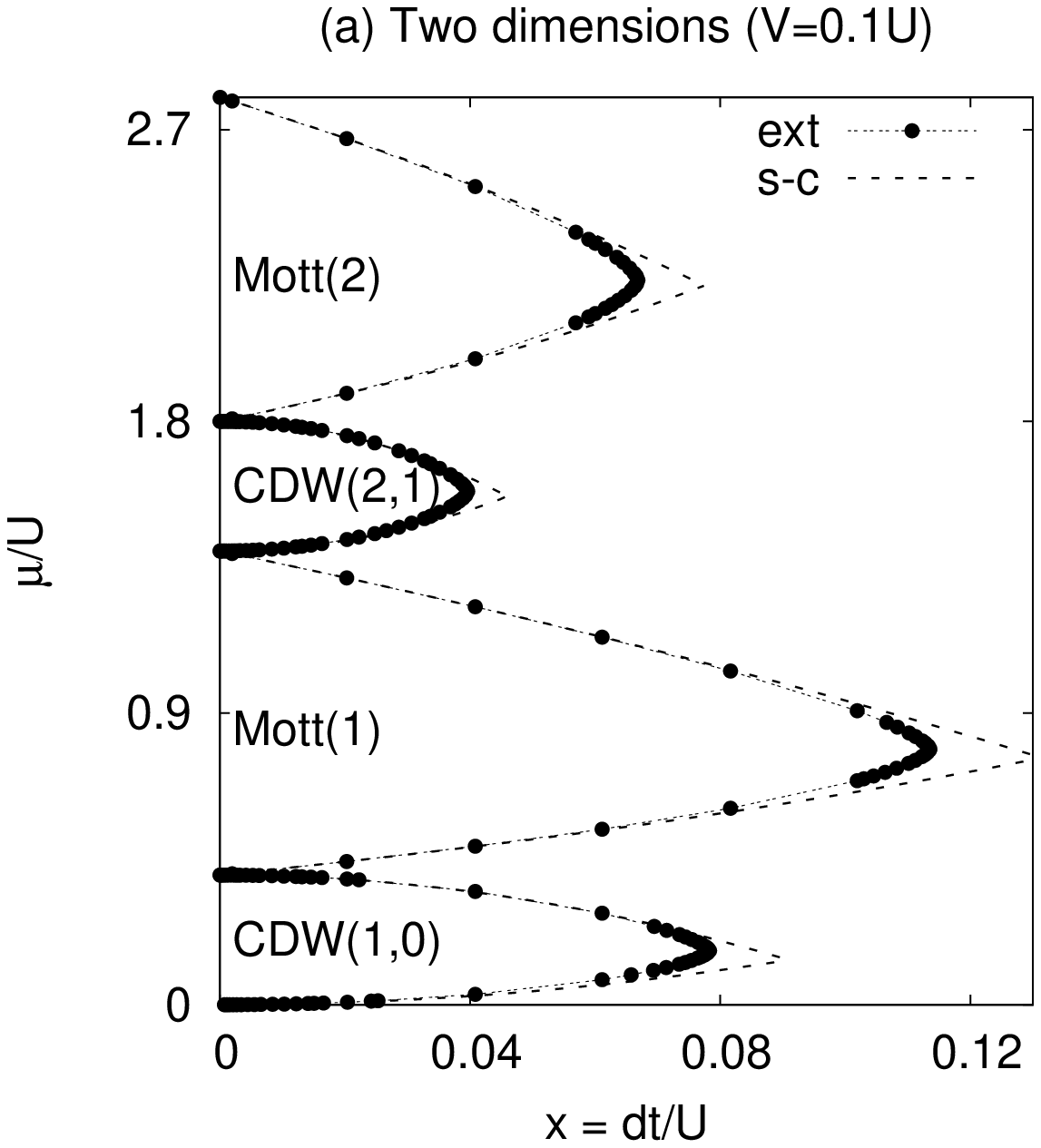}}}
\centerline{\scalebox{0.6}{\includegraphics{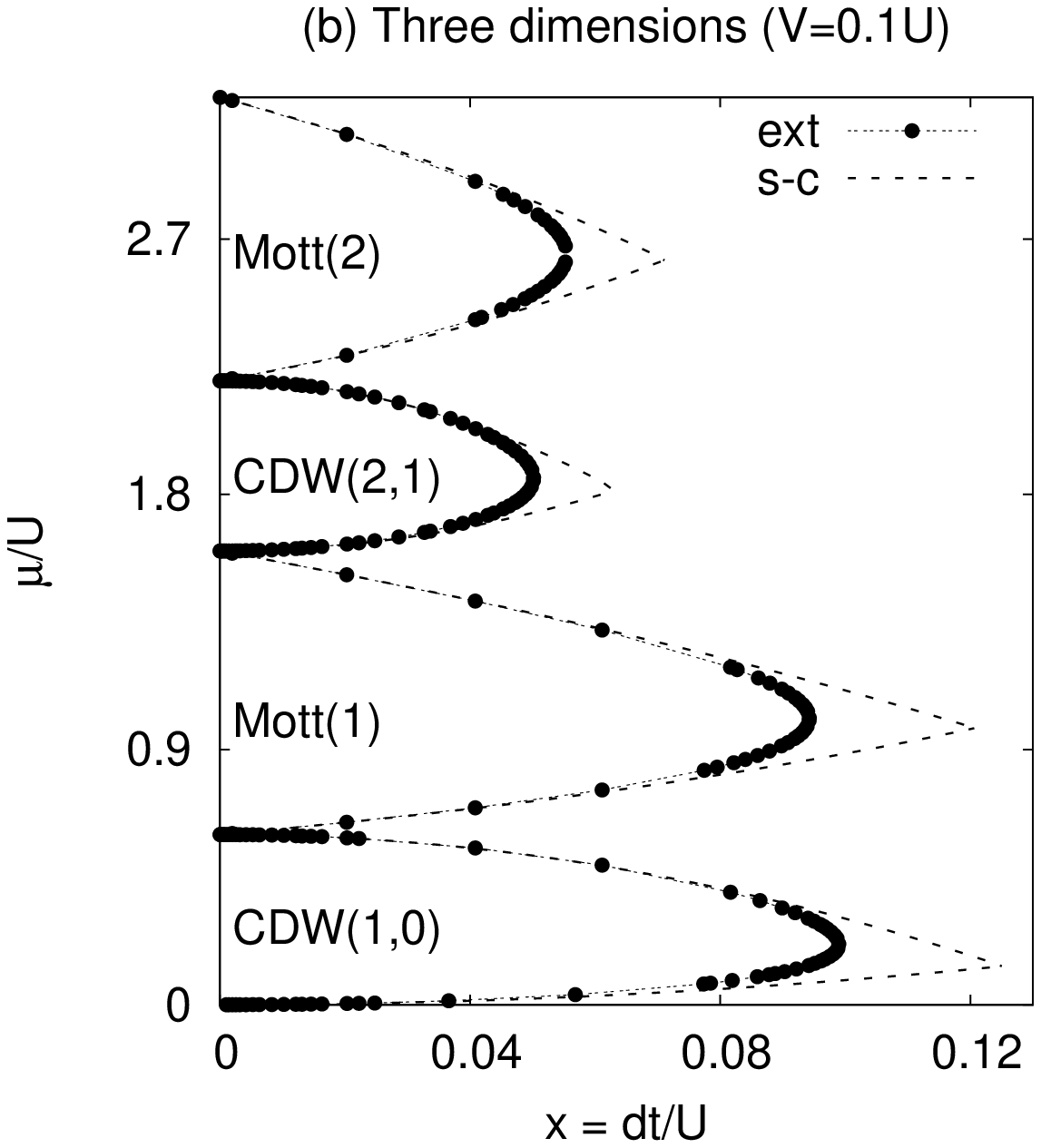}}}
\caption{\label{fig:pd}
We show the chemical potential $\mu$ (in units of $U$) versus $x = dt/U$ 
phase diagram for (a) two- ($d = 2$) and (b) three- ($d=3$) dimensional 
hypercubic lattices. We choose the nearest-neighbor repulsion as $V = 0.1U$. 
The dotted lines correspond to phase boundaries for the Mott insulator to 
superfluid and CDW insulator to supersolid states as determined from the 
third-order strong-coupling perturbation theory (s-c). The circles 
correspond to the extrapolation fit (ext) discussed in the text.
}
\end{figure}
\subsection{Extrapolation to Infinite order via Scaling Theory}
\label{sec:ext}

As a general rule, the third-order strong-coupling perturbation theory appears
to be more accurate in lower dimensions. For this reason, an extrapolation 
technique to infinite order in $t$ is highly desirable to determine more 
accurate phase diagrams. Here, we propose a chemical potential extrapolation 
technique based on scaling theory to extrapolate our third-order power-series 
expansion into a functional form that is appropriate for the Mott and CDW lobes. 

It is known that the critical point at the tip of the Mott and CDW lobes 
has the scaling behavior of a ($d+1$)-dimensional $XY$ model, and therefore 
the lobes have Kosterlitz-Thouless shapes for $d=1$ and power-law shapes 
for $d > 1$. For the latter case considered in this manuscript, 
we propose the following ansatz for the Mott and CDW lobes which includes 
the known power-law critical behavior of the tip of the lobes
\begin{align}
\label{eqn:smu}
\frac{\mu_{\rm Mott/CDW}^{\rm par,hol}}{U} &= A_{\rm Mott/CDW}(x) \nonumber \\
&\pm B_{\rm Mott/CDW}(x)(x_{\rm Mott/CDW}^c-x)^{z\nu},
\end{align}
where 
$
A_{\rm Mott/CDW}(x) = a_{\rm Mott/CDW} + b_{\rm Mott/CDW} x 
+ c_{\rm Mott/CDW} x^2 + d_{\rm Mott/CDW} x^3 + ...
$ 
and 
$
B_{\rm Mott/CDW}(x) = \alpha_{\rm Mott/CDW} + \beta_{\rm Mott/CDW} x 
+ \gamma_{\rm Mott/CDW} x^2 + \delta_{\rm Mott/CDW} x^3 + ...
$ 
are regular functions of $x = d t/U$, 
$x_{\rm Mott/CDW}^c$ is the critical point which determines the location of
the Mott and CDW lobes, 
and $z\nu$ is the critical exponent for the ($d+1$)-dimensional $XY$ model
which determines the shape of the Mott and CDW lobes near $x_{\rm Mott/CDW}^c$.
In Eq.~(\ref{eqn:smu}), the plus sign corresponds to the particle branch, and
the minus sign corresponds to the hole branch.
The parameters $a_{\rm Mott/CDW}$, $b_{\rm Mott/CDW}$, $c_{\rm Mott/CDW}$ and $d_{\rm Mott/CDW}$ 
depend on $U$, $V$ and $n_0$ or $\{ n_a, n_b \}$, and they are determined by 
matching them with the coefficients given by our third-order expansion such that
$
A_{\rm Mott/CDW}(x) = (\mu_{\rm Mott/CDW}^{\rm par} + \mu_{\rm Mott/CDW}^{\rm hol})/2.
$
To determine the $U$, $V$ and $n_0$ or $\{ n_a, n_b \}$ dependence of the 
parameters $\alpha_{\rm Mott/CDW}$, $\beta_{\rm Mott/CDW}$, 
$\gamma_{\rm Mott/CDW}$, $\delta_{\rm Mott/CDW}$, $x_{\rm Mott/CDW}^c$ and $z\nu$, 
we first expand the left hand side of 
$
B_{\rm Mott/CDW}(x)(x_{\rm Mott/CDW}^c-x)^{z\nu} = (\mu_{\rm Mott/CDW}^{\rm par} - \mu_{\rm Mott/CDW}^{\rm hol})/2
$
in powers of $x$, and match the coefficients with the coefficients given by our
third-order expansion. Then we fix $z\nu$ at its well-known values such that 
$z\nu \approx 2/3$ for $d = 2$ and $z\nu = 1/2$ for $d > 2$, and 
set $\delta_{\rm Mott/CDW} = 0$ to determine $\alpha_{\rm Mott/CDW}$, 
$\beta_{\rm Mott/CDW}$, $\gamma_{\rm Mott/CDW}$ and $x_{\rm Mott/CDW}^c$ 
self-consistently.

Having discussed the strong coupling perturbation theory, next we present the
ground-state phase diagrams for $(d=2)$- and ($d = 3$)-dimensional hypercubic 
lattices.

\subsection{Numerical Results}
\label{sec:numeric}

In Fig.~\ref{fig:pd}, the results of the third-order strong-coupling perturbation 
theory (dotted lines) are compared to those of the extrapolation technique 
(circles) when $V = 0.1U$. 
At $t = 0$, the chemical potential width of all Mott and CDW lobes are $U$ 
and $0.1z U$, respectively where $z = 2d$, and that the ground state alternates 
between the CDW and Mott phases as a function of $\mu$. 
For instance, the ground state is a vacuum 
$(n_0=0) $ for $\mu \le 0$; it is a CDW with
$(n_a=1, n_b=0)$ for $0 \le \mu \le 0.1zU$; it is a Mott insulator with $(n_0=1)$ 
for $0.1zU \le \mu \le (1+0.1z)U$; it is a CDW with $(n_a=2, n_b=1)$ 
for $(1+0.1z)U \le \mu \le (1+0.2z)U$; it is a Mott insulator with $(n_0=2)$ for 
$(1+0.2z)U \le \mu \le (2+0.2z)U$.

\begin{figure} [htb]
\centerline{\scalebox{0.6}{\includegraphics{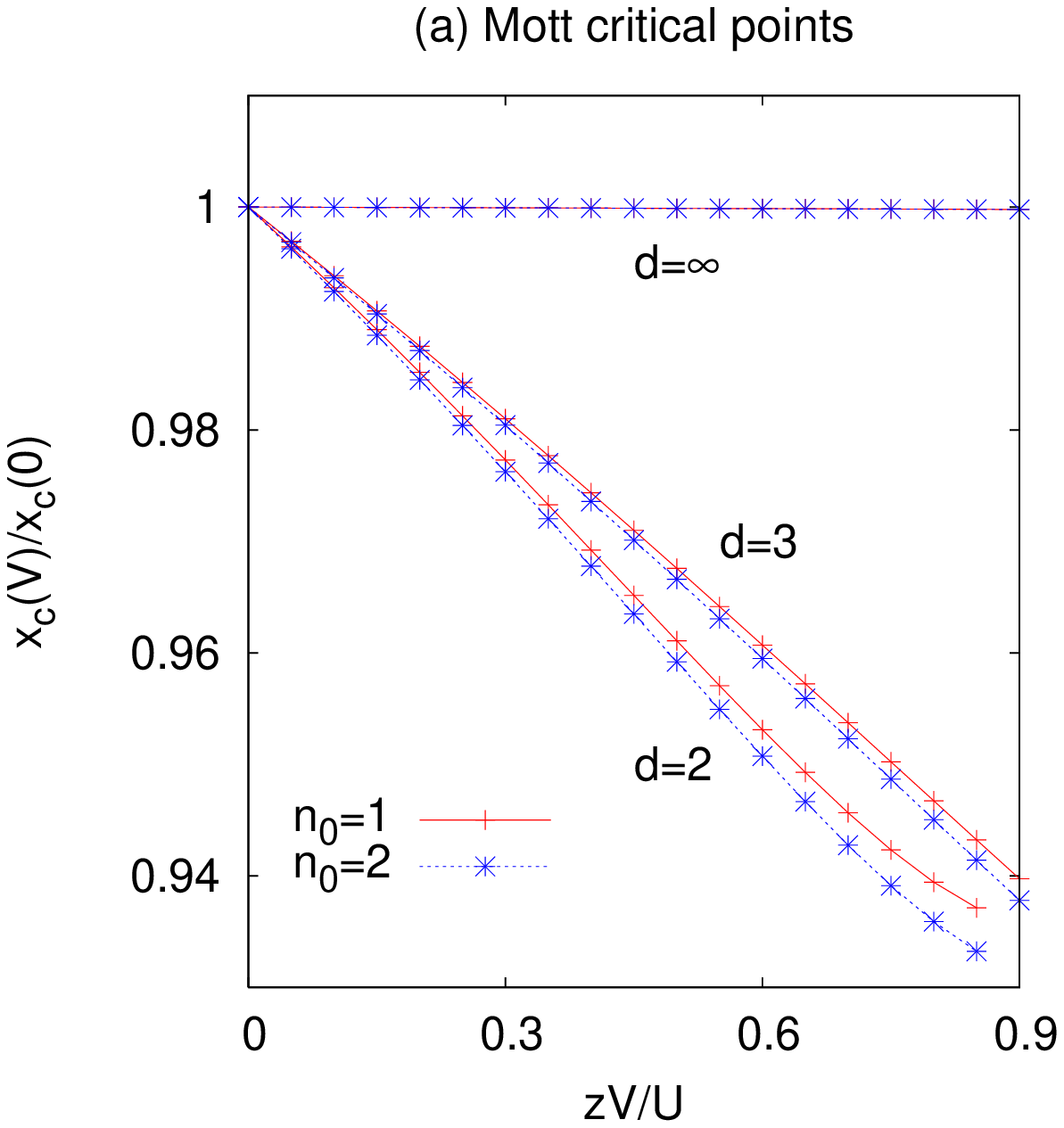}}}
\centerline{\scalebox{0.6}{\includegraphics{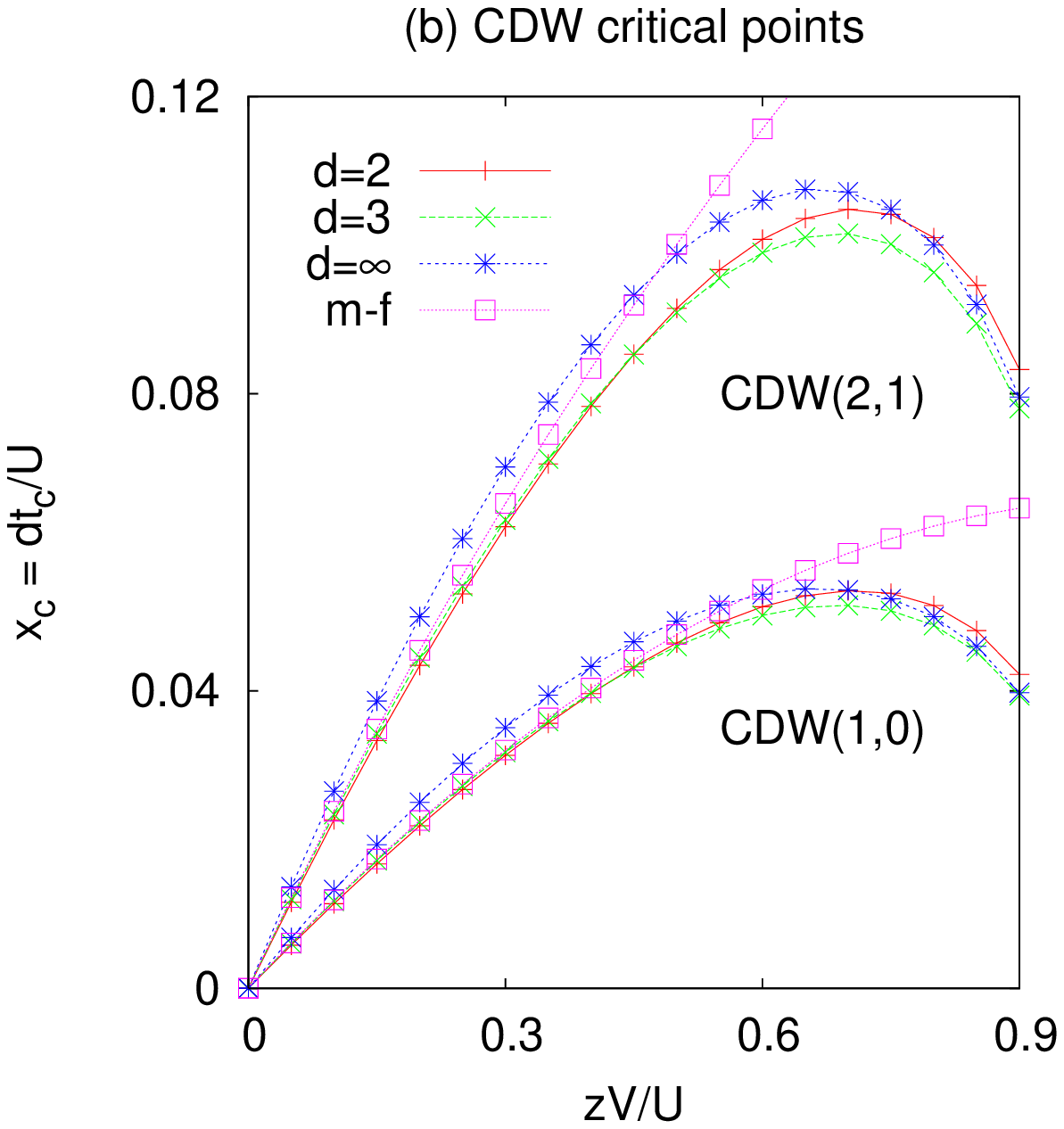}}}
\caption{\label{fig:cp} (Color online)
We show the critical points (location of the tips) $x_c = dt_c/U$ that are found 
from the chemical potential extrapolation technique described in the text 
versus $zV/U$, where $z = 2d$. In Fig.~(a), $x_c$'s are scaled with their $V = 0$ value; 
in infinite dimensions the exact critical hoppings for the Mott lobes are independent of $V$.
In Fig.~(b), comparing the extrapolated strong-coupling and exact mean-field 
results for the $d \to \infty$ limit shows that the critical points 
for the CDW lobes become less accurate as $V$ increases. This is because 
the coefficient of the $O(t^4)$ term in the power series becomes very large 
when $z V \approx 0.7U$, which also causes an unphysical decrease in $x_c$ 
for $zV \gtrsim 0.7U$ after an initial increase.
}
\end{figure}

As $t$ increases from zero, the range of $\mu$ about which the ground state 
is a Mott insulator or CDW decreases, and the Mott insulator and CDW phases disappear at 
a critical value of $t$, beyond which the system becomes a superfluid 
near the Mott lobes or a supersolid near the CDW lobes.
In addition, similar to what was found for the on-site BH model~\cite{freericks-1, freericks-2},
the strong-coupling expansion 
overestimates the phase boundaries, and it leads to unphysical pointed 
tips for all Mott and CDW lobes. This is not surprising since a finite-order 
perturbation theory cannot describe the physics of the tricritical point correctly. 

In Fig.~\ref{fig:cp}, we show the critical points (location of the tips) 
$x_c = dt_c/U$ versus $zV/U$. In Fig.~\ref{fig:cp}(a), $x_c$'s of the Mott 
lobes are scaled with their $V = 0$ value.
The critical points are calculated with the chemical potential extrapolation 
technique that is based on the scaling theory with the exponent $z\nu$ fixed to 
its known value. It is expected that the locations
of the tips of the CDW lobes to increase as a function of $V$, because the 
presence of a nonzero $V$ is what allowed these states to form in the first 
place (the Mott insulator critical points tend to move in as $V$ increases). 
Comparing the extrapolated strong-coupling and exact mean-field (to be discussed below)
results for the $d \to \infty$ limit shows that the critical points 
for the CDW lobes become less accurate as $V$ increases.
It turns out that the coefficient of the $O(t^4)$ term in the 
power series is generally small for the Mott lobes, but it can become very 
large for the CDW lobes when $z V \sim U$.
We remind that we assume $U > zV$ in this manuscript. 
As shown in Fig.~\ref{fig:cp}(b), This also causes an unphysical decrease in $x_c$ 
for $zV \gtrsim 0.7U$ after an initial increase. Therefore, inclusion 
of the $O(t^4)$ terms in the expansion are necessary to improve the accuracy 
of the phase boundaries near the tips of the CDW lobes when $z V \sim U$.
In addition, we present a short list of $V/U$ versus the critical points 
$x_c = dt_c/U$ in Table~\ref{table:x_c} for $(d = 2)$- and $(d = 3)$-dimensional 
lattices. 

As a further check of the accuracy of our perturbative expansion, next
we compare $d \to \infty$ limit of our results to the mean-field one 
which corresponds to the exact solution on an ($d \to \infty$)-dimensional 
hypercubic lattice.

\begin{center}
\begin{table*} [htb]
\caption{\label{table:x_c} 
We list the critical points (location of the tips) $x_c = dt_c/U$ that are found from
the chemical potential extrapolation technique described in the text. 
}
\begin{tabular}{c|cccc|cccc}
\hline \hline
& & Two dimensions & & & & Three dimensions & & \\
\cline{2-9}
$V/U$ & CDW(1,0) & Mott(1) & CDW(2,1) & Mott(2) & CDW(1,0) & Mott(1) & CDW(2,1) & Mott(2) \\
\hline
0.00  & -        & 0.117 & -       & 0.0691   & -       & 0.0981 & -      & 0.0578  \\
0.01	& 0.00929 & 0.117 & 0.00465 & 0.0689  	& 0.0143	& 0.0977 & 0.00717 & 0.0576 \\
0.02	& 0.0183   & 0.116 & 0.00916 & 0.0687  	& 0.0278	& 0.0974 & 0.0139	& 0.0574	\\
0.03	& 0.0270   & 0.116 & 0.0135	 & 0.0684 	& 0.0405	& 0.0970 & 0.0203	& 0.0571	\\
0.04	& 0.0354   & 0.116 & 0.0178	 & 0.0682 	& 0.0522	& 0.0966 & 0.0263	& 0.0569	\\
0.05	& 0.0434   & 0.115 & 0.0219	 & 0.0680 	& 0.0630	& 0.0962 & 0.0317	& 0.0567	\\
0.06  & 0.0512   & 0.115 & 0.0258  & 0.0678   & 0.0723  & 0.0958 & 0.0367 & 0.0564  \\
0.07	& 0.0586   & 0.115 & 0.0295	 & 0.0676 	& 0.0814	& 0.0955 & 0.0411	& 0.0562	\\
0.08  & 0.0656   & 0.114 & 0.0331  & 0.0673   & 0.0888  & 0.0951 & 0.0449 & 0.0559  \\
0.09  & 0.0721   & 0.114 & 0.0365  & 0.0671   & 0.0947  & 0.0947 & 0.0480 & 0.0557  \\
0.10	& 0.0783   & 0.114 & 0.0396	 & 0.0669 	& 0.0990	& 0.0942 & 0.0502	& 0.0555	\\
\hline \hline
\end{tabular}
\end{table*}
\end{center}
\section{Mean-Field Decoupling Theory}
\label{sec:mf}

In the large-dimensional case, mean-field theory becomes exact, 
so examining the mean-field theory for the extended BH model provides 
another way to validate the strong-coupling expansion and to test to 
see how well the scaling result produces the correct phase diagram.

In constructing the mean-field theory, one first defines the superfluid order parameter 
as $\varphi_k = \langle b_k \rangle$ where $\langle ... \rangle$ 
is the thermal average, and then replaces the operator $b_k$ with 
$\varphi_k + \delta b_k$ in the hopping term of Eq.~(\ref{eqn:ebhh}). 
This approximation decouples the two-particle hopping term into 
single-particle ones, and the resultant mean-field Hamiltonian can be 
solved via exact diagonalization in a power series of $\varphi_k$.
The order parameter is finite ($\varphi_k \ne 0$) for the superfluid 
and supersolid ground states, and it vanishes ($\varphi_k = 0$) for 
the Mott and CDW phases. Therefore, $\varphi_k \to 0^+$ signals the 
phase boundary between an incompressible and a compressible phase. 
The generalized order parameter equation to the case of $V \ne 0$ 
can be written as~\cite{menotti}
\begin{equation}
\label{eqn:mf-op}
\varphi_k = \bar{\varphi}_k t \left[\frac{n_k+1}{U n_k + V_k^{\rm dip} -\mu} 
- \frac{n_k}{U (n_k-1) + V_k^{\rm dip} - \mu} \right],
\end{equation}
where $\bar{\varphi}_k = \sum_{{\langle k' \rangle}_k} \varphi_{k'}$ is the 
sum of the order parameters at sites $k'$ neighboring to site $k$, 
and $V_k^{\rm dip} = V \sum_{{\langle k' \rangle}_k} n_{k'}$ is the interaction 
of one atom with sites $k'$ neighboring to the site $k$. 

To determine the phase boundary between the Mott and superfluid phases
from Eq.~(\ref{eqn:mf-op}), we set $\varphi_k = \varphi_0$, $\bar{\varphi}_k = z\varphi_0$, 
and $V_k^{\rm dip} = z V n_0$. Since $\varphi_0 \to 0^+$ near the phase 
boundary, Eq.~(\ref{eqn:mf-op}) can be satisfied only if
\begin{equation}
\label{eqn:mf-mott}
\frac{1}{z t} = \frac{n_0+1}{U n_0 + z V n_0 -\mu} -\frac{n_0}{U (n_0-1) + z V n_0 - \mu},
\end{equation}
which gives a quadratic equation for $\mu$. Notice that this equation 
recovers the known result for the on-site BH model when 
$V = 0$~\cite{fisher, stoof}, and it can be easily solved to obtain
\begin{align}
\label{eqn:mf-mu}
\mu_{\rm Mott}^{\rm par,hol} &= U(n_0-1/2) + z V n_0 - z t/2 \nonumber \\
& \pm \sqrt{U^2/4 - U(n_0+1/2)z t + z^2 t^2},
\end{align}
where the plus sign corresponds to the particle branch, and the minus sign 
corresponds to the hole branch. In the $d \to \infty$ limit, we checked that 
our strong-coupling perturbation results for the Mott lobes agree with this exact 
solution when the latter is expanded out to third order in $t$, providing an 
independent check of the algebra (one must note that the terms $V$ and $2V$
that appear in the denominator vanish in the limit when $d \to \infty$ because
$V \propto 1/d$).
Equation~(\ref{eqn:mf-mu}) also shows that 
the Mott lobes are separated by $z V$, but their shapes are independent of $V$; 
in particular, the critical points for the Mott lobes are independent of $V$.

To determine the phase boundary between the CDW and supersolid phases from
Eq.~(\ref{eqn:mf-op}), we set $\varphi_i = \varphi_A$, $\bar{\varphi}_i = z\varphi_B$
and $V_i^{\rm dip} = z V n_B$ for $i \in$ A sublattice, and we set $\varphi_j = \varphi_B$, 
$\bar{\varphi}_j = z\varphi_A$ and $V_j^{\rm dip} = z V n_A$ for $j \in$ B sublattice.
This leads to two coupled equations for $\varphi_A$ and $\varphi_B$. 
Since $\{\varphi_A, \varphi_B\} \to 0^+$ near the phase boundary, 
Eq.~(\ref{eqn:mf-op}) can be satisfied only if 
\begin{align}
\label{eqn:mf-cdw}
\frac{1}{z^2 t^2} &= \left[ \frac{n_a+1}{U n_a + z V n_b - \mu} 
- \frac{n_a}{U (n_a-1) + z V n_b - \mu} \right] \nonumber \\
&\left[ \frac{n_b+1}{U n_b + z V n_a -\mu} -\frac{n_b}{U (n_b-1) + z V n_a -\mu} \right],
\end{align}
which gives a quartic equation for $\mu$. Since a simple closed form 
analytic solution for $\mu$ is not possible, we solve Eq.~(\ref{eqn:mf-cdw}) 
with Mathematica for each of the CDW lobes separately. 
In the $d \to \infty$ limit, we also checked 
that our strong-coupling perturbation results for the CDW lobes agree 
with this exact solution when the latter is expanded out to third order 
in $t$, providing again an independent check of the algebra.

\begin{figure} [htb]
\centerline{\scalebox{0.6}{\includegraphics{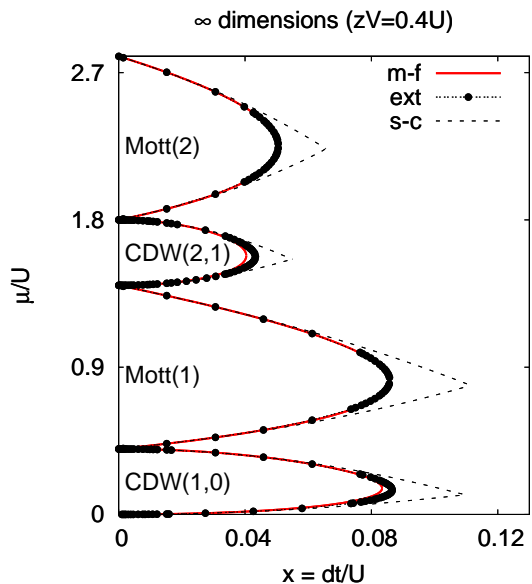}}}
\caption{\label{fig:mf} (Color online)
We show the chemical potential $\mu$ (in units of $U$) versus $x = dt/U$ 
phase diagram for a ($d \to \infty$)-dimensional hypercubic lattice. 
Here the nearest-neighbor repulsion scales inversely with $d$ 
such that $zV = 0.4U$. The dotted lines correspond to phase 
boundaries for the Mott insulator to superfluid and CDW insulator to supersolid 
states as determined from the third-order strong-coupling perturbation theory (s-c). 
The circles correspond to the extrapolation fit (ext) discussed in the text. 
The red solid lines correspond to phase boundaries for the Mott insulator 
to superfluid and CDW insulator to supersolid states as determined from 
the mean-field theory (m-f) which becomes exact for $d \to \infty$.
}
\end{figure}

In Fig.~\ref{fig:mf}, the results of the third-order strong-coupling 
perturbation theory (dotted lines) is compared to those of the exact 
mean-field theory (red solid lines) and of the extrapolation technique 
(circles) for an infinite ($d \to \infty$)-dimensional hypercubic lattice when 
$zV = 0.4U$. Notice that, in infinite dimensions, both $t$ and $V$ must 
scale inversely with $d$ such that $d t$ and $d V$ are finite. 
The extrapolated solutions are indistinguishable from the exact ones 
for the Mott lobes, and they are within $5\%$ of each 
other for the tips of the CDW lobes. It turns out that this minor disagreement around
the tips of the CDW lobes is due to the large coefficient of the $O(t^4)$ term in 
the power-series expansion. Therefore, we conclude that, even in infinite 
dimensions, the agreement of the third-order strong-coupling perturbation theory 
with the exact mean-field theory is quite good.

\section{Conclusions}
\label{sec:conclusions}

We analyzed the zero temperature phase diagram of the extended Bose-Hubbard (BH)
model with on-site and nearest-neighbor boson-boson repulsions in 
($d > 1$)-dimensional hypercubic lattices.
We used the many-body version of Rayleigh-Schr\"odinger perturbation theory 
in the kinetic energy term with respect to the ground state of the system 
when the kinetic energy term is absent.
This technique was previously used to discuss the phase diagram of the
on-site BH model~\cite{freericks-1, freericks-2}, and its extrapolated results showed an excellent 
agreement with the recent Quantum Monte Carlo simulations~\cite{prokofiev-1,prokofiev-2}. 
Here, we generalized this method to the extended BH model, hoping 
to develop an analytical approach which could be as accurate as the numerical ones. 

We derived analytical expressions for the phase boundaries between the 
incompressible (Mott or charge-density-wave (CDW) insulators) and 
compressible (superfluid or supersolid) phases up to third order in the hopping $t$. 
However, we remark that the strong-coupling perturbation theory developed 
here cannot be used to calculate the phase boundary between two 
compressible phases, \textit{e.g.} the supersolid to superfluid transition.
We also proposed a chemical potential extrapolation technique 
based on the scaling theory to extrapolate our third-order power
series expansion into a functional form that is appropriate for the Mott 
or CDW lobes.

We believe some of our results could potentially be observed with 
ultracold dipolar Bose gases loaded into optical lattices~\cite{goral, menotti}. 
This is motivated by the recent success in observing superfluid to 
Mott insulator transition with ultracold point-like Bose gases loaded 
into optical lattices. Such lattices are created by the intersection 
of laser fields, and they are nondissipative periodic potential 
energy surfaces for the atoms. An ultracold dipolar Bose gas can be
realized in many ways with optical lattices. For instance, heteronuclear 
molecules which have permanent electric dipole moments, 
Rydberg atoms which have very large induced electric dipole moment, 
or Chromium-like atoms which have large intrinsic magnetic moment, 
etc. can be used to generate sufficiently strong long-ranged 
dipole-dipole interactions.

This work can be extended in several ways if desired. For instance, our current 
results for the CDW phase are not directly applicable to the one-dimensional 
case. We are currently working on this problem and will report our results 
elsewhere. In addition, it turns out that the coefficient of the $O(t^4)$ 
term in the power series is generally small for the Mott lobes, 
but it can become very large for the CDW lobes when $z V \sim U$. 
Therefore, inclusion of the $O(t^4)$ is necessary to improve the accuracy 
of the phase boundaries near the tips of the CDW lobes 
when $z V \sim U$. Lastly, one can include the next-nearest-neighbor 
repulsion term to the current model, which would lead to 
additional CDW phases.  One can also examine how the momentum 
distribution changes with the hopping in the CDW phase, or in the 
Mott phase when there is a nearest-neighbor repulsion. 
This last calculation could have direct relevance for experiments 
on these systems and would generalize recent results for the 
$V=0$ case~\cite{freericks-3}.

\section{Acknowledgements}
\label{sec:ack}

We would like to thank E. Tiesinga for many useful discussions. 
J. K. F. acknowledge support under ARO Grant W911NF0710576 with
funds from the DARPA OLE Program.


\begin{thebibliography}{99}
\bibitem{jaksch} D. Jaksch, C. Bruder, J. I. Cirac, C. W. Gardiner, and P. Zoller, Phys. Rev. Lett. \textbf{81}, 3108 (1998).
\bibitem{greiner} M. Greiner, O. Mandel, T. Esslinger, T.W. H\"ansch, and I. Bloch,  Nature (London), \textbf{415}, 39 (2002).
\bibitem{spielman-1} I. B. Spielman, W. D. Phillips, and J. V. Porto, Phys. Rev. Lett. \textbf{98}, 080404 (2007).
\bibitem{spielman-2} I. B. Spielman, W. D. Phillips, and J. V. Porto, Phys. Rev. Lett. \textbf{100}, 120402 (2008).
\bibitem{bloch} F. Gerbier, S. Trotzky, S. Fölling, U. Schnorrberger, J. D. Thompson, A. Widera, I. Bloch, L. Pollet, M. Troyer, B. Capogrosso-Sansone, N. V. Prokof’ev, and B. V. Svistunov, Phys. Rev. Lett. {\bf 101}, 155303 (2008).
\bibitem{fisher} M. P. A. Fisher, P. B. Weichman, G. Grinstein, and D. S. Fisher, Phys. Rev. B \textbf{40}, 546 (1989).
\bibitem{freericks-1} J. K. Freericks and H. Monien, Europhys. Lett. \textbf{24}, 545 (1994).
\bibitem{freericks-2} J. K. Freericks and H. Monien, Phys. Rev. B \textbf{53}, 2691 (1996).
\bibitem{ye-1} S. Ospelkaus, A. Pe'er, K.-K. Ni, J. J. Zirbel, B. Neyenhuis, S. Kotochigova, P. S. Julienne, J. Ye, and D. S. Jin, Nature Physics \textbf{4}, 622 (2008).
\bibitem{ye-2} K.-K. Ni, S. Ospelkaus, M. H. G. de Miranda, A. Pe'er, B. Neyenhuis, J. J. Zirbel, S. Kotochigova, P. S. Julienne, D. S. Jin, and J. Ye, Science \textbf{322}, 231 (2008).
\bibitem{bruder} C. Bruder, Rosario Fazio, and Gerd Sch\"on, Phys. Rev. B \textbf{47}, 342 (1993).
\bibitem{parhat} Parhat Niyaz, R. T. Scalettar, C. Y. Fong, and G. G. Batrouni, Phys. Rev. B \textbf{50}, 362  (1994).
\bibitem{otterlo} Anne van Otterlo, Karl-Heinz Wagenblast, Reinhard Baltin, C. Bruder, Rosario Fazio, and Gerd Sch\"on, Phys. Rev. B \textbf{52}, 16176 (1995).
\bibitem{kuhner} Till D. K\"uhner, Steven R. White, and H. Monien, Phys. Rev. B \textbf{61}, 12474 (2000).
\bibitem{kovrizhin} D. L. Kovrizhin, G. Venketeswara Pai, and S. Sinha, Europhys. Lett. \textbf{72}, 162 (2005).
\bibitem{leggett} A. J. Leggett, Phys. Rev. Lett. \textbf{25}, 1543 (1970).
\bibitem{chan} E. Kim and M. H. W. Chan, Science \textbf{305}, 1941 (2004).
\bibitem{landau} L. D. Landau and L. M. Lifshitz, \textit{Quantum Mechanics}, Butterworth-Heinemann (1981).
\bibitem{prokofiev-1} B. Capogrosso-Sansone, N. V. Prokof'ev, and B. V. Svistunov, Phys. Rev. B \textbf{75}, 134302   (2007).
\bibitem{prokofiev-2} B. Capogrosso-Sansone, S. G. S\"oyler, N. Prokof'ev, and B. Svistunov, Phys. Rev. A \textbf{77}, 015602 (2008).
\bibitem{menotti} C. Trefzger, C. Menotti, and M. Lewenstein, Phys. Rev. A \textbf{78}, 043604 (2008).
\bibitem{stoof} D. van Oosten, P. van der Straten, and H. T. Stoof, Phys. Rev. A \textbf{63}, 053601 (2001).
\bibitem{goral} K. Goral, L. Santos, and M. Lewenstein, Phys. Rev. Lett. \textbf{88}, 170406 (2002).
\bibitem{freericks-3} J. K. Freericks, H. R. Krishnamurthy, Yasuyuki Kato, Naoki Kawashima, and Nandini Trivedi, preprint, arXiv:0902.3435 (2009).

\end{thebibliography}
\end{document}